\documentstyle[12pt]{article}

\begin{document}

\makeatletter
\makeatother


\begin{center}
{\large{\bf Spin Operator and Entanglement in Quantum Field Theory}}
\end{center}
\vskip .5 truecm
\centerline{\bf  Kazuo Fujikawa$^1$, C.H. Oh$^{2}$ and Chengjie Zhang$^{2}$}
\vskip .4 truecm
\centerline {\it $^1$ Mathematical Physics Laboratory,}
\centerline {\it RIKEN Nishina Center, Wako 351-0198, Japan}
\vspace{0.3cm}
\centerline {\it $^2$ Center for Quantum Technologies,}
\centerline {\it  National University of Singapore, Singapore 117543, Singapore}
\vskip 0.5 truecm

\begin{abstract}
Entanglement is studied in the framework of Dyson's S-matrix theory in relativistic quantum field theory, which leads to a natural definition of entangled states of a particle-antiparticle pair and the spin operator from a Noether current. As an explicit example, the decay of a massive pseudo-scalar particle into a pair of electron and positron is analyzed. Two spin operators are extracted from the Noether current. The Wigner spin operator characterizes spin states at the rest frame of each fermion and, although not measurable in the laboratory, gives rise to a straightforward generalization of low energy analysis of entanglement to the ultra-relativistic domain. In contrast, if one adopts a (modified) Dirac spin operator, the entanglement measured by spin correlation becomes maximal near the threshold of the decay, while the entanglement is replaced by the classical correlation for the ultra-relativistic electron-positron pair by analogy to the case of neutrinos, for which a hidden-variables-type description is possible.
 Chiral symmetry differentiates the spin angular momentum and the magnetic moment.
 The use of weak interaction that can measure helicity is suggested in the analysis of entanglement at high energies  instead of a Stern-Gerlach apparatus which is useless for the electron. A difference between the electron spin at high energies and the photon linear polarization is also noted. The Standard Model can describe all the observable properties of leptons.
\end{abstract}



\section{Introduction}
Starting with the analysis of Einstein, Podolsky and Rosen~\cite{epr}, the entanglement in non-relativistic quantum mechanics has been much discussed, in particular, in connection with hidden-variables models~\cite{bell, chsh, cirel'son}. Also, the basic issues such as non-locality and causality have been deliberated. It is however clear that the issues related to relativity are better treated in a fully relativistic formulation. There exists ample literature on this subject, but we just mention some of them~\cite{pryce, fleming,czachor,anderson, terno, alsing, terashima, pachos, caban1, caban2, friis, choi} where further references are found. We are mainly interested in the basic issues such as   the natural definition of entangled states and  spin operators, the possible energy dependence of entanglement measured by spin correlation, and non-locality and causality.
We formulate the entanglement in the framework of relativistic quantum field theory, to be precise, in the S-matrix theory defined by Dyson~\cite{dyson2}. In S-matrix theory, we treat only asymptotic states which contain particles far apart from each other. As a concrete example, we present such a formulation by studying the decay of a massive pseudo-scalar particle into a pair of electron and positron~\cite{particledata}. The decay of a pseudo-scalar particle  itself has been discussed in~\cite{caban1}, but we present a fully field theoretical formulation and emphasize  many different aspects of the problem including experimental measurements. Our formulation is very simple and straightforward and we recognize several novel features of quantum mechanical entanglement.  

To be specific, we show novel aspects related to, for example, the derivation of two kinds of spin operators, where one of them is directly measurable at the laboratory and the other is not, starting with the Noether current, and the energy-scale dependence of entanglement of two-spin systems depending on the choice of spin operators thus constructed. Quite independently of a specific choice of the spin operator, an entangled state is constructed by a local and causal interaction in the S-matrix theory in a manner perfectly consistent with the uncertainty principle. Other novel features are the crucial role of chiral symmetry, which is not clearly recognized in the first quantization, in the analysis of spin related freedom of fermions, in particular, to distinguish  spin operators from the magnetic moment, and a unified treatment of electrons and neutrinos.  We also discuss the use of parity violating weak interactions to test the spin entanglement of high energy electrons and clarify the difference in the energy dependence of the electron spin and the linear polarization of the photon. We present a view that the Standard Model can describe all the observable properties of leptons.

\section{Massive pseudo-scalar particle decay}

We  consider the decay of a very massive pseudo-scalar particle $P$ into a pair of electron and positron described by the free Lagrangian (in the natural units $c=\hbar=1$)~\cite{bjorken},
\begin{eqnarray}
{\cal L}_{0}=\bar{\psi}(x)[i\gamma^{\mu}\partial_{\mu}-m]\psi(x) + \frac{1}{2}[\partial_{\mu}P(x)\partial^{\mu}P(x)-M^{2}P(x)^{2}]
\end{eqnarray}
and an (effective) interaction Lagrangian or Hamiltonian, 
\begin{eqnarray}
H_{I}(t)=g\int d^{3}x: P(x)\bar{\psi}(x)i\gamma_{5}\psi(x):
\end{eqnarray}
with a coupling constant $g$. The Dyson formula for the S-matrix~\cite{dyson2, bjorken, weinberg},
$S=1-i\int dt H_{I}(t)+ ....$,
shows that the final state of the decay of the heavy particle $P$ at rest is described by $\Psi=S|P(\vec{0})\rangle$, namely, by ignoring the forward amplitude,
\begin{eqnarray}
\Psi =-ig\int d^{4}x: P(x)\bar{\psi}(x)i\gamma_{5}\psi(x):|P(\vec{0})\rangle,
\end{eqnarray}
where we assume a small $g$.
We expand the free fields in the interaction picture
\begin{eqnarray}
\psi(x)&=&\int\frac{d^{3}p}{(2\pi)^{3/2}}\sum_{s}[a(\vec{p},s)u(\vec{p},s)e^{-ipx}+b^{\dagger}(\vec{p},s)v(\vec{p},s)e^{ipx}],\nonumber\\
P(x)&=&\int\frac{d^{3}p}{(2\pi)^{3/2}}\sqrt{\frac{1}{2E}}[c(\vec{p})e^{-ipx}+c^{\dagger}(\vec{p})e^{ipx}]
\end{eqnarray}
where the solutions of the free Dirac equation are given by~\cite{bjorken} 
\begin{eqnarray}
&&u(\vec{p},s)=\sqrt{\frac{E+m}{2E}}\left(\begin{array}{c}
            \xi(s)\\
            \frac{\vec{\sigma}\cdot\vec{p}}{E+m}\xi(s)
            \end{array}\right), \ \ \
v(\vec{p},s)=\sqrt{\frac{E+m}{2E}}\left(\begin{array}{c}
            \frac{\vec{\sigma}\cdot\vec{p}}{E+m}\xi(-s)\\
            \xi(-s)
            \end{array}\right)
\end{eqnarray}
with $\vec{\sigma}$ standing for Pauli matrices.
The two-component spinor $\xi(\pm s)$ for a fixed unit spin vector $\vec{s}$ (at the rest frame of the electron or positron) satisfies
\begin{eqnarray}
\vec{s}\cdot\vec{\sigma}\xi(\pm s)=\pm \xi(\pm s),\ \
\xi^{\dagger}(s)\xi(s^{\prime})=\delta_{s,s^{\prime}},
\end{eqnarray}
and the expressions (5) are obtained by Lorentz boosting to the direction of momentum.
For example, 
for $\vec{s}=\vec{p}/|\vec{p}|$, $s$ agrees with  helicity eigenvalues $h\xi(\pm s)\equiv(\vec{p}\cdot\vec{\sigma})\xi(\pm s)/|\vec{p}|=\pm \xi(\pm s)$, which is actually twice of the conventional definition of helicity. The commutation relations in the continuum notation are
$[c({\vec{p}}^{\ \prime}), c^{\dagger}(\vec{p})]=\delta^{3}({\vec{p}}^{\ \prime}-\vec{p})$ and $\{a({\vec{p}}^{\ \prime},s^{\prime}),a^{\dagger}(\vec{p},s)\}=\{b({\vec{p}}^{\ \prime},s^{\prime}),b^{\dagger}(\vec{p},s)\}=\delta_{s,s^{\prime}}\delta^{3}({\vec{p}}^{\ \prime}-\vec{p})$, and all others are vanishing.
 
The final asymptotic state (3) has the structure (using $|P(\vec{0})\rangle=c^{\dagger}(\vec{0})|0\rangle$)
\begin{eqnarray}
\Psi
=\frac{g}{\sqrt{4\pi M}}\int d^{3}pd^{3}p^{\prime}\delta^{4}(p+p^{\prime}-P)\sum_{s,s^{\prime}}[\bar{u}(\vec{p},s)\gamma_{5}v(-\vec{p},s^{\prime})]a^{\dagger}(\vec{p},s)b^{\dagger}(-\vec{p},s^{\prime})|0\rangle,
\end{eqnarray}
with $P=(\vec{0}, M)$.
We thus define for the fixed momentum direction of the electron, 
\begin{eqnarray}
\Psi(\vec{p})&\equiv&\frac{1}{\sqrt{2}}
[a^{\dagger}(\vec{p},s)b^{\dagger}(-\vec{p},-s)+a^{\dagger}(\vec{p},-s)b^{\dagger}(-\vec{p},s)]|0\rangle,
\end{eqnarray}
which is valid for any choice of the spin vector $\vec{s}$ by noting $\bar{u}(\vec{p},s)\gamma_{5}v(-\vec{p},s^{\prime})=\delta_{s,-s^{\prime}}$.  (Actually,  $s=1$ in the expression of (8) but we keep $s$ to indicate the direction of $\vec{s}$). This shows a way to prepare a desired state in the framework of local and causal relativistic field theory. All the properties of the asymptotic state (8) are accounted for in the framework which is consistent with locality, causality and the uncertainty principle, as is discussed further later; in particular, it is important to recognize that we integrate over the entire Minkowski space in (3), namely, we have no information about when and where the particle decayed. In passing, we mention that a scalar particle decay, instead of a pseudo-scalar, is
also naturally described by field theory and the result is obtained by replacing $\bar{u}(\vec{p},s)\gamma_{5}v(-\vec{p},s^{\prime})$ by $\bar{u}(\vec{p},s)v(-\vec{p},s^{\prime})$ in (7), but the final result is more involved due to the opposite intrinsic parity of a fermion and an anti-fermion.  

In the following analysis, it is convenient to imagine a very large 3-dimensional box with a volume $L^{3}$ and impose periodic boundary conditions so that we have discretized momentum $\vec{p}$ and $\{a(\vec{p},s), a^{\dagger}(\vec{p}^{\ \prime},s^{\prime})\}=\delta{s,s^{\prime}}\delta_{\vec{p},\vec{p}^{\ \prime}}(L/2\pi)^{3}$, for example, and the replacement $\int d^{3}p\rightarrow (2\pi/L)^{3}\sum_{\vec{p}}$. Various formulas become simple if one sets $2\pi/L\equiv m_{L}=1$ by choosing a suitable unit of mass; for example, $\langle 0|a(\vec{p},s) a^{\dagger}(\vec{p},s^{\prime}) |0\rangle=\delta{s,s^{\prime}}$.
Note that we have the relation $2\sqrt{\vec{p}^{2}+m^{2}}=2E=M$ in (8). 
\\

\noindent{\bf Noether charge}\\

The conserved angular momentum operator (Noether charge) which generates the rotational symmetry of the Dirac action is given by~\footnote{It is confirmed that
the free Dirac action $S=\int d^{4}x \bar{\psi}(x)[i\gamma^{\mu}\partial_{\mu}-m]\psi(x)$ is invariant under an infinitesimal global rotation of the field $\psi^{\prime}(x)=\exp\{-i\vec{\omega}[\vec{L}+\vec{S}]\}\psi(x)$ with an infinitesimal  constant $\vec{\omega}$. By making $\vec{\omega}$ time-dependent $\vec{\omega}(t)$, one obtains $S^{\prime}=\int d^{4}x \bar{\psi}^{\prime}(x)[i\gamma^{\mu}\partial_{\mu}-m]\psi^{\prime}(x)=S+\int dt\int d^{3}x \partial_{0}\vec{\omega}(t)\psi^{\dagger}(x)[\vec{L}+\vec{S}]\psi(x)$ which defines the Noether charge $\vec{J}=\int d^{3}x\psi^{\dagger}(x)[\vec{L}+\vec{S}]\psi(x)$. This is the most general procedure to define the Noether current (and charge) in path integral. See, for example, K. Fujikawa and H. Suzuki, {\em Path Integrals and Quantum Anomalies}, (Oxford University Press, 2004). }   
\begin{eqnarray}
\hat{\vec{J}}&=&\int d^{3}x:\psi^{\dagger}(x)[\vec{L}+\vec{S}]\psi(x):
\end{eqnarray}
with the orbital part $\vec{L}=\vec{x}\times(-i\vec{\nabla})=\vec{p}\times(-i\vec{\nabla}_{p})$ and the spin part
\begin{eqnarray}
\vec{S}=\frac{1}{2}\left(\begin{array}{cc}
            \vec{\sigma}& 0\\
            0&\vec{\sigma}
            \end{array}\right).
\end{eqnarray}
The  angular momentum operator $\hat{\vec{J}}$ generates the transformation
\begin{eqnarray}
[\hat{\vec{J}}, \psi(x)]=[\vec{L}+\vec{S}]\psi(x),
\end{eqnarray}
and it is  written as  
\begin{eqnarray}
\hat{\vec{J}}&=&\int d^{3}p\sum_{s,s^{\prime}}\{[u^{\dagger}(\vec{p},s^{\prime})\left(\vec{L}+\vec{S}\right)u(\vec{p},s)]a^{\dagger}(\vec{p},s^{\prime})a(\vec{p},s)\nonumber\\
&&-[v^{\dagger}(\vec{p},s)\left(\vec{L}+\vec{S}\right)v(\vec{p},s^{\prime})]b^{\dagger}(\vec{p},s^{\prime})b(\vec{p},s)\}\nonumber\\
&+&\int d^{3}p\sum_{s}\{a^{\dagger}(\vec{p},s)\left(\vec{L}a(\vec{p},s)\right)+b^{\dagger}(\vec{p},s)\left(\vec{L}b(\vec{p},s)\right)\}.
\end{eqnarray}
It is natural to define the first term in (12) as spin part of the angular momentum operator $\hat{\vec{J}}_{S}$, which is re-written as
\begin{eqnarray}
\hat{\vec{J}}_{S}&\equiv&\int d^{3}p\sum_{s,s^{\prime}}\{[u^{\dagger}(\vec{p},s^{\prime})\left(\vec{L}+\vec{S}\right)u(\vec{p},s)]a^{\dagger}(\vec{p},s^{\prime})a(\vec{p},s)\nonumber\\
&&-[v^{\dagger}(\vec{p},s)\left(\vec{L}+\vec{S}\right)v(\vec{p},s^{\prime})]b^{\dagger}(\vec{p},s^{\prime})b(\vec{p},s)\}\nonumber\\
&=&\int d^{3}p\sum_{s,s^{\prime}}\frac{1}{2}\{\xi(s^{\prime})^{\dagger}\vec{\sigma}\xi(s)a^{\dagger}(\vec{p},s^{\prime})a(\vec{p},s)-\xi^{\dagger}(-s)\vec{\sigma}\xi(-s^{\prime})b^{\dagger}(\vec{p},s^{\prime})b(\vec{p},s)\},
\end{eqnarray}
and it depends only on $\vec{\sigma}$.
The second term in (12) may be defined as orbital part of the angular momentum operator $\hat{\vec{J}}_{L}$. 
Using this expression of $\hat{\vec{J}}_{S}$, one can confirm $[\hat{\vec{J}}_{L},\hat{\vec{J}}_{S}]=0$ and the $SO(3)$ (to be precise $SU(2)$) algebras $[\hat{J}^{a}_{S},\hat{J}^{b}_{S}]=i\epsilon^{abc}\hat{J}^{c}_{S}$ and $[\hat{J}^{a},\hat{J}^{b}]=i\epsilon^{abc}\hat{J}^{c}$ which constrain the eigenvalues of $\hat{J}^{a}_{S}$ and  $\hat{J}^{b}$ to be half odd-integers. 
 
 One can confirm $\hat{\vec{J}}_{S}\Psi(\vec{p})=0$ for the state in (8) with fixed momentum, namely, $\hat{\vec{J}}_{S}$ does not rotate the momentum direction and the state $\Psi(\vec{p})$ is a singlet under the spin rotation. The amplitude and the spin part of the angular momentum operator are the direct generalizations of common non-relativistic expressions for the singlet state 
$\psi_{{\rm singlet}}= |+\rangle|-\rangle-|-\rangle|+\rangle$
for any chosen direction $\vec{s}$ of the spin vector. The operator $\hat{\vec{J}}_{S}$ in (13) generates the group $SO(3)$, which is basically the spin rotation at the {\em rest frame} of each particle (Wigner spin in~\cite{terno}), for any value of momentum independently of the magnitude of mass. The operator $\hat{\vec{J}}_{S}$ in (13) characterizes the {\em spin states} of the fermion, and all the familiar analyses of entanglement at low energies are formally  generalized to the ultra-relativistic domain if one adopts the spin operator $\hat{\vec{J}}_{S}$. Note, however, that the expectation value of spin  in the laboratory frame 
\begin{eqnarray}
\langle 0|a(\vec{p},s)\hat{\vec{J}}_{S}a^{\dagger}(\vec{p},s)|0\rangle=(1/2)\xi(s)^{\dagger}\vec{\sigma}\xi(s), 
\end{eqnarray}
holds independently of the value of electron momentum $\vec{p}$ for any direction of the spin  vector $\vec{s}$ defined at its rest frame. Experimentally, this property is not associated with the spin operator; see the spin angular momentum in (A.5) in~\cite{tolhoek}, and references therein.

 The operator $\hat{\vec{J}}_{S}$ in (13) does not explicitly incorporate the fact that the proper part of spin rotation is violated by fixed $\vec{p}$ in the Dirac equation in the laboratory frame. It may also be desirable to define the notion of spin by incorporating Wigner's little group,  $SO(3)$ (to be precise $SU(2)$) for a massive particle and $E(2)$ for a massless particle~\cite{weinberg}, and possibly a smooth transition of the (effective) little group  for the electron from $SO(3)$ to $E(2)$ depending on its energy~\cite{coester} in a massive particle decay. The Lorentz factor we have in mind is of the order  $m/E\simeq 1/140$ for the neutral pion decay and $m/E\sim 10^{-5}$ for a possible pseudo-scalar Higgs-like particle  by recalling $E=\sqrt{\vec{p}^{2}+m^{2}}=M/2$ in our problem. 
\\

\noindent{\bf Modified Dirac spin operator}\\

We thus  examine an alternative definition of  the spin operator using the proper spin part of the angular momentum operator $\hat{\vec{J}}_{S}(\vec{p})=\hat{\vec{S}}(\vec{p})+\hat{\vec{L}}(\vec{p})$ in (13), namely, 
\begin{eqnarray}
\hat{\vec{S}}(\vec{p})&\equiv&\sum_{s,s^{\prime}}\{[u^{\dagger}(\vec{p},s^{\prime})\vec{S}u(\vec{p},s)]a^{\dagger}(\vec{p},s^{\prime})a(\vec{p},s)\nonumber\\
&&-[v^{\dagger}(-\vec{p},s)\vec{S}v(-\vec{p},s^{\prime})]b^{\dagger}(-\vec{p},s^{\prime})b(-\vec{p},s)\}\nonumber\\
&=&\sum_{s,s^{\prime}}\{[\frac{1}{2}\frac{m}{E}\xi^{\dagger}(s^{\prime})\vec{\sigma}_{T}\xi(s)+\frac{1}{2}\hat{p}\xi^{\dagger}(s^{\prime})(\vec{\sigma}\cdot\hat{p})\xi(s)]a^{\dagger}(\vec{p},s^{\prime})a(\vec{p},s)\\
&&-[\frac{1}{2}\frac{m}{E}\xi^{\dagger}(-s)\vec{\sigma}_{T}\xi(-s^{\prime})+\frac{1}{2}\hat{p}\xi^{\dagger}(-s)(\vec{\sigma}\cdot\hat{p})\xi(-s^{\prime})]b^{\dagger}(-\vec{p},s^{\prime})b(-\vec{p},s)\},\nonumber
\end{eqnarray}
where we consider for simplicity the fixed momentum sector which is relevant to our problem; $\hat{p}$ stands for a unit vector in the direction of $\vec{p}$ and $\vec{\sigma}_{T}$ stands for the transverse components. The use of this spin operator is analogous to the use of the transverse field components of the photon; it is not manifestly Lorentz (and rotation) invariant by itself, but for the given invariant expression (12) at any frame one can uniquely identify the spin operator. (Note that $\int d^{3}x:\psi^{\dagger}(x)\vec{S}\psi(x):$ in (9), which is referred to as "Dirac spin operator" in~\cite{terno}, is not time-independent by itself. Our construction is to first define the time-independent momentum space expression (13) and then extract $\hat{\vec{S}}(\vec{p})$ in (15). Our operator is thus a {\em variant} of the Dirac spin operator, and we tentatively call it "modified Dirac spin operator".)  Our proposal in (15), which defines the spin of the moving electron observed at the laboratory frame, is related to the classical construction using the center of mass coordinate~\cite{pryce, fleming} and the use of the Pauli-Lubanski vector in the first quantization~\cite{czachor}, while $\hat{\vec{J}}_{S}$ in (13) is related to the definition of the spin operator used in the first quantization~\cite{anderson,terno,caban1}. 

We start with an analysis near the threshold of the decay $\vec{p}\simeq 0$ 
($M\simeq 2m$), namely, for the extremely non-relativistic electron and positron for which the momentum is negligibly small compared to the rest mass. We then have
\begin{eqnarray}
\hat{\vec{S}}(\vec{0})=\sum_{s,s^{\prime}}\frac{1}{2}\{\xi(s^{\prime})^{\dagger}\vec{\sigma}\xi(s)a^{\dagger}(\vec{0},s^{\prime})a(\vec{0},s)-\xi^{\dagger}(-s)\vec{\sigma}\xi(-s^{\prime})b^{\dagger}(\vec{0},s^{\prime})b(\vec{0},s)\},
\end{eqnarray}
which agrees with the angular momentum operator $\hat{\vec{J}}_{S}(\vec{0})(=\hat{\vec{J}}(\vec{0}))$ and the asymptotic state becomes a non-relativistic one 
\begin{eqnarray}
\Psi(\vec{0})=\frac{1}{\sqrt{2}}[a^{\dagger}(\vec{0},s)b^{\dagger}(\vec{0},-s)+a^{\dagger}(\vec{0},-s)b^{\dagger}(\vec{0},s)]|0\rangle,
\end{eqnarray}
where $s$ stands for the spin component specified by the spin vector $\vec{s}$ at the rest frame of the electron or positron which coincides with the center of mass and also the laboratory frame. Namely, the amplitude and spin operator are the same as the common non-relativistic expressions for a singlet state 
$\psi_{{\rm singlet}}= |+\rangle|-\rangle-|-\rangle|+\rangle$
for {\em any} chosen direction $\vec{s}$ of the spin. The relative plus sign for a singlet state in (17) is confirmed to be consistent by operating  $\hat{\vec{S}}(\vec{0})$ on the state $\hat{\vec{S}}(\vec{0})\Psi(\vec{0})=0$.

We now analyze the non-negligible momentum sector.
We first note  the relation
\begin{eqnarray}
u^{\dagger}(\vec{p},s^{\prime})S^{m}u(\vec{p},s)=\frac{1}{2}\bar{u}(\vec{p},s^{\prime})\gamma^{m}\gamma_{5}u(\vec{p},s)
\end{eqnarray}
in our convention.
Namely, our spin operator is formally regarded as an axial-vector quantity under the boost, and the enhancement of the longitudinal component relative to transverse components in (15) is an analogue of the Lorentz "contraction"; the time component vanishes at the rest frame $\bar{u}(\vec{0},s^{\prime})\gamma^{0}\gamma_{5}u(\vec{0},s)=0$, and thus the longitudinal component is minimum at the rest frame if one remembers 
the Lorentz invariant quantity $(\bar{u}\gamma^{m}\gamma_{5}u)^{2}-(\bar{u}\gamma^{0}\gamma_{5}u)^{2}$ after correcting the factor $m/E$~\cite{bjorken}.  

The longitudinal component parallel to the momentum $\hat{p}$ in $\hat{\vec{S}}(\vec{p})$
agrees with the longitudinal component of the angular momentum operator (13) and it gives the helicity generated by $\vec{\sigma}\cdot\hat{p}$ which is also a generator of the little group $E(2)$ of a massless particle. For the helicity basis, the longitudinal component  
$\hat{h}=\vec{p}\cdot\hat{\vec{S}}(\vec{p})/|\vec{p}|=\vec{p}\cdot\hat{\vec{J}}(\vec{p})/|\vec{p}|$ gives the helicity operator,
\begin{eqnarray}
\hat{h}(\vec{p})=\frac{1}{2}\sum_{h=\pm}[h a^{\dagger}(\vec{p},h)a(\vec{p},h)- h b^{\dagger}(-\vec{p},h)b(-\vec{p},h)],
\end{eqnarray}
for the asymptotic state (8) written as
\begin{eqnarray}
\Psi(\vec{p})=\frac{1}{\sqrt{2}}[a^{\dagger}(\vec{p},+)b^{\dagger}(-\vec{p},+)+a^{\dagger}(\vec{p},-)b^{\dagger}(-\vec{p},-)]|0\rangle
\end{eqnarray}
where $\pm$ stand for the (twice of) helicity $h\equiv 2\times(1/2)\vec{p}\cdot\vec{\sigma}/|\vec{p}|$. We note that the combination $\sum_{s}a(\vec{p},s)\xi(s)$, for example, has an invariant meaning with respect to the choice of the spin vector $\vec{s}$, and in particular, the relations
$\sum_{s}a(\vec{p},s)\xi(s)=\sum_{h}a(\vec{p},h)\xi(h)$  and 
 $\sum_{s}b^{\dagger}(-\vec{p},s)\xi(-s)=\sum_{h}b^{\dagger}(-\vec{p},h)\xi(-h)$ hold.   

The expression (15) shows that the measured value $\langle \vec{e}\cdot\hat{\vec{S}}(\vec{p})\rangle$ is half odd-integer ($\pm 1/2$) for $\vec{e}=\vec{s}=\vec{p}/|\vec{p}|$ (of which direction is common to the laboratory frame and the electron or positron rest frame) for the helicity eigenstate such as $a^{\dagger}(\vec{p},h)|0\rangle$, i.e., {\em dispersion-free} by taking only eigenvalues, while the measured value of $\langle \vec{e}\cdot\hat{\vec{S}}(\vec{p})\rangle$ for $\vec{e}=\vec{s}\perp \vec{p}$ (which is also common to the laboratory frame and the electron or positron rest frame)
is away from half odd-integer ($\pm 1/2$)  for the state such as $a^{\dagger}(\vec{p},s)|0\rangle$, i.e.,
{\em dispersion-full} by taking values away from eigenvalues for $\vec{p}\neq 0$. This expresses the fact that the spin components perpendicular to the momentum cannot be diagonalized simultaneously with the Hamiltonian for a fixed $\vec{p}\neq 0$ in the Dirac equation, although the (rotation invariant) energy eigenvalues are degenerate with respect to spin freedom. Our spin operator corresponds to the average of the spin angular momentum operator in the first quantization with respect to (momentum space) spinor solutions at each Lorentz frame. 
 In particular, for the extremely relativistic electron or positron with $|\vec{p}|\rightarrow \infty$ ($M/m\rightarrow\infty$), the transverse component of the spin operator vanishes, $\vec{e}\cdot\hat{\vec{S}}(\vec{p})\rightarrow 0$, for the vector $\vec{e}=\vec{s}\perp\vec{p}$. This means that the electron or positron state approaches the {\em chirality} (and helicity) eigenstates defined by projection operators $(1\pm \gamma_{5})/2$ and thus the transverse spin component of a free fermion is completely indeterminate, $1/2$ or $-1/2$ randomly, in each measurement at the laboratory; this also means that the helicity flip is suppressed for the ultra-relativistic case.
  The correlation between the spin index of $a(\vec{p},s)^{\dagger}$ appearing in the state vector (8) and the measured spin  at the laboratory is completely lost for $\vec{e}=\vec{s}\perp\vec{p}$ at extreme high energies. 
\\

\noindent {\bf Magnetic moment}
\\  

We now mention the Pauli-type coupling  
\begin{eqnarray}
(\mu/2)\int d^{3}x\bar{\psi}F_{\mu\nu}\sigma^{\mu\nu}\psi
\end{eqnarray}
to the constant background magnetic field $\vec{B}$ (by writing only the particle-number conserving part),
\begin{eqnarray}
&&\mu\vec{B}\cdot\int d^{3}p\sum_{s,s^{\prime}}\{[\frac{1}{2}\xi(s^{\prime})^{\dagger}\vec{\sigma}_{T}\xi(s)+
\frac{1}{2}\frac{m}{E}\hat{p}\xi(s^{\prime})^{\dagger}(\vec{\sigma}\cdot\hat{p})\xi(s)]a^{\dagger}(\vec{p},s^{\prime})a(\vec{p},s)\nonumber\\
&&+[\frac{1}{2}\xi(-s)^{\dagger}\vec{\sigma}_{T}\xi(-s^{\prime})+
\frac{1}{2}\frac{m}{E}\hat{p}\xi(-s)^{\dagger}(\vec{\sigma}\cdot\hat{p})\xi(-s^{\prime})]b^{\dagger}(\vec{p},s^{\prime})b(\vec{p},s)\},
\end{eqnarray}
which shows the enhancement of the transverse spin components $\vec{\sigma}_{T}$ for large $|\vec{p}|$ in contrast to (15) since we here deal with $\bar{\psi}\vec{S}\psi$ instead of $\psi^{\dagger}\vec{S}\psi$, and these two operators have different chiral properties. In a review article of electron polarization~\cite{tolhoek}, $\psi^{\dagger}\vec{S}\psi$ is called the "spin angular momentum" and  $\bar{\psi}\vec{S}\psi$ is called the "magnetic moment". The Pauli coupling breaks chiral symmetry, which is not recognized in the first quantization. For the relatively weak transverse magnetic field, the Pauli coupling induces a small energy splitting $\sim\pm\mu B$ which in turn leads to the chirality (and helicity) flip of the longitudinally polarized energetic fermion~\cite{fujikawa}; note that a helicity eigenstate is always represented as a linear superposition of transversely  polarized states inside a transverse magnetic field. Inside the strong transverse magnetic field, the electron is no more free and tends to be polarized in the direction of $\vec{B}$.
For example, inside a high energy synchrotron, electrons eventually become transversely polarized by emitting radiation; such a state will be represented by a superposition of helicity eigenstates, which is no more the eigenstate of chirality.  The magnetic interaction breaks chiral symmetry.

We thus have 3 basic operators to describe the spin freedom of fermions; Wigner spin operator (13), the modified Dirac spin operator (15) and the magnetic moment (22). 

\section{Entanglement and correlation}
We here argue that the asymptotic state (17) characterized by $\hat{\vec{S}}(\vec{0})$ is maximally {\em entangled} in accord with conventional analysis, but the asymptotic state (20) characterized by the helicity operator (19) with vanishing 
transverse component $\vec{e}\cdot\hat{\vec{S}}(\vec{p})\rightarrow 0$ for $\vec{e}\perp \vec{p}$ and $|\vec{p}| \rightarrow \infty$ is classically {\em correlated}. 

One has a continuous number of spin projectors 
\begin{eqnarray}
&&P_{a}(\pm s^{\prime})=|a(\vec{0},\pm s^{\prime})\rangle\langle a(\vec{0},\pm s^{\prime})|\otimes 1,\nonumber\\
&&P_{b}(\pm s^{\prime\prime})=1\otimes |b(\vec{0},\pm s^{\prime\prime})\rangle\langle b(\vec{0},\pm s^{\prime\prime})|,
\end{eqnarray}
in (17) for a fixed non-relativistic momentum $\vec{p}\simeq 0$ with the  arbitrary  directions of $\vec{s}^{\ \prime}$ and $\vec{s}^{\ \prime\prime}$. Here we used a simplified notation $P_{a}(\pm s)=|a(\vec{0},\pm s)\rangle\langle a(\vec{0},\pm s)|\otimes 1$ for $P_{a}(\pm s)=a^{\dagger}(\vec{0},\pm s)|0\rangle\langle 0|a(\vec{0},\pm s)\otimes 1$, for example. It is known that the non-contextual (and local) hidden-variables representation which reproduces all the properties of quantum mechanics is not possible in this system, (17) and (23), with the dimension of the Hilbert space $d=2\times 2=4$~\cite{gleason, kochen, beltrametti}. The inseparable asymptotic state (17) is thus quantum mechanically entangled with respect to spin variables, in agreement with the conventional analysis~\footnote{The possible non-locality associated with entanglement is usually discussed in the context of local realism represented by hidden-variables models~\cite{bell, chsh}. 
We here follow Gisin in understanding the CHSH inequality~\cite{chsh}, namely, it gives a necessary and sufficient separability condition of pure quantum mechanical states without direct reference to non-locality (although no interactions among two parties are assumed)~\cite{werner, gisin}; in this proof of Gisin's theorem  it is important to take all the possible combinations of projection operators into consideration. Our analysis here is performed in this spirit.}.  The $d=4$ hidden-variables model used by Bell in his original paper~\cite{bell} is classified as non-contextual and local (i.e., applied to only far apart systems)~\cite{mermin}, and thus his model and his inequality cannot describe entanglement.  See Eq.(30) later for a more explicit analysis.

On the other hand, for the asymptotic state (20) with a {\em fixed} direction of $\vec{p}$ and $|\vec{p}| \rightarrow \infty$, we have a very limited number of effective spin projection operators available, since the helicity flip is suppressed in (15), and we have only  
\begin{eqnarray}
&&P_{a}(\pm)=|a(\vec{p},\pm)\rangle\langle a(\vec{p},\pm)|\otimes 1,\nonumber\\
&&P_{b}(\pm)=1\otimes |b(\vec{p},\pm)\rangle\langle b(\vec{p},\pm)|,
\end{eqnarray}
with $P_{a}(+)+P_{a}(-)=1, \ P_{a}(+)P_{a}(-)=0$ and $P_{b}(+)+P_{b}(-)=1, \ P_{b}(+)P_{b}(-)=0$. Here $\pm$ stand for helicity indices. (One may recall the neutrinos which are described by their helicity, even though the neutrinos are considered to have small but non-vanishing masses~\cite{particledata}.) The actual observables are given by the helicity operator in (19), for example, 
\begin{eqnarray}
\langle P_{a}(+)\hat{h}(\vec{p})_{a}P_{a}(+)P_{b}(+)\hat{h}(\vec{p})_{b}P_{b}(+)\rangle.
\end{eqnarray}
It is thus possible to have a consistent non-contextual and local hidden-variables-type representation for (24)~\cite{beltrametti}:
One may define classical variables $U_{a}(\psi,\lambda)$ and $D_{a}(\psi,\lambda)$ 
corresponding to $P_{a}(+)$ and $P_{a}(-)$, respectively, with 
\begin{eqnarray}
U_{a}(\psi,\lambda)+D_{a}(\psi,\lambda)=1,\ \ U_{a}(\psi,\lambda)D_{a}(\psi,\lambda)=0
\end{eqnarray}
where $U_{a}(\psi,\lambda)$ and $D_{a}(\psi,\lambda)$ assume the eigenvalues of $1$ or $0$ of projection operators depending on the hidden variables $\lambda$. Similarly one may define
$U_{b}(\psi,\lambda)$ and $D_{b}(\psi,\lambda)$, and may impose the subsidiary conditions 
$U_{a}(\psi,\lambda)=U_{b}(\psi,\lambda)$ and $D_{a}(\psi,\lambda)=D_{b}(\psi,\lambda)$. One can then confirm that it is possible to reproduce all the correlations with fixed $\vec{p}$, by choosing suitable functions $U$ and $D$ for the specific state in (20) and helicity operators in (19), for example,
\begin{eqnarray}
&&\langle P_{a}(-)P_{b}(+)\rangle=\int d\lambda \rho(\lambda)D_{a}(\psi,\lambda)U_{b}(\psi,\lambda)=0,\nonumber\\
&&\frac{\langle P_{a}(+)P_{b}(+)\rangle}{\langle P_{b}(+)\rangle}=\frac{\int d\lambda \rho(\lambda)U_{a}(\psi,\lambda)U_{b}(\psi,\lambda)}{\int d\lambda \rho(\lambda)U_{b}(\psi,\lambda)}=1,
\end{eqnarray}
using  a suitable non-negative normalized weight function $\rho(\lambda)$, 
together with the relations
\begin{eqnarray}
&&\int d\lambda \rho(\lambda)U_{a}(\psi,\lambda)=\int d\lambda \rho(\lambda)D_{a}(\psi,\lambda)=1/2,\nonumber\\
&&\int d\lambda \rho(\lambda)U_{b}(\psi,\lambda) = \int d\lambda \rho(\lambda)D_{b}(\psi,\lambda)=1/2
\end{eqnarray}
implied by 
$\langle P_{a}(+)\rangle=\langle P_{a}(-)\rangle=\langle P_{b}(+)\rangle
=\langle P_{b}(-)\rangle=1/2$. These relations show that the spin (helicity) correlation in the limit $|\vec{p}|\rightarrow\infty$ is consistently described by non-contextual and local hidden-variables models and thus "classically correlated", although the asymptotic state (20) itself is not separable. 
This situation is analogous to the case where one can measure only the electric charges but not spin polarizations of the electron-positron pair in the decay, and the use of the term "classical correlation" may be natural.

If one uses the modified Dirac spin operator (15), the {\em entanglement measured by spin correlation} becomes a notion which depends on the energy scale involved (or the boost of two particles in opposite directions) in relativistic quantum field theory, at least in the present specific example of spin $1/2$ particles. This may imply that fermion spin measured at the laboratory is not an ideal means to describe entanglement. In contrast, if one adopts the spin operator defined in (13) (Wigner spin) for the asymptotic state in (8), which characterizes the spin {\em states} at the rest frame of each particle and thus such spins of two moving particles {\em are not simultaneously measured} at any fixed Lorentz frame, the entanglement (as a theoretical characterization of quantum states) has an invariant meaning independently of the energy scale. The energy dependence of entanglement measured by fermion spin correlation as we have argued, which is related to Lorentz contraction, is somewhat analogous to the energy dependence of the lifetime of an unstable particle; the lifetime of a particle has an invariant meaning at its rest frame, but the measured lifetime in the laboratory depends on its velocity.

\section{Discussion and conclusion}

As for the experimental test of the helicity structure (20) with (19) and suppressed helicity flip, one may  in principle consider the weak interaction mediated by charged currents, $e\rightarrow \nu$ or $e^{+}\rightarrow \bar{\nu}$, for which only the left-handed currents take part in the interaction~\cite{weinberg}. Note the {\em unification} of electro and weak interactions at high energies.
Our interaction Hamiltonian (2) is written as
\begin{eqnarray}
H_{I}(t)=g\int d^{3}x P(x)i[\bar{\psi}_{L}(x)\psi_{R}(x)-\bar{\psi}_{R}(x)\psi_{L}(x)],
\end{eqnarray}
and thus the chirality $\psi_{R,L}=[(1\pm \gamma_{5})/2]\psi$ is always correlated, but it becomes a good quantum number only for the  massless fermion. 
For the extremely relativistic fermion, the chirality is identified with helicity, and the left-handed fermion corresponds to a particle with helicity $h=-(1/2)$ or an anti-particle with  helicity $h=+(1/2)$. Thus if one confirms the charged current weak interaction for the electron in (20), the partner 
positron  also has $h=-(1/2)$ (i.e., right-handed) and thus {\em no} charged current weak interaction. If one applies the strong transverse magnetic field, one may observe the effect of spin rotation, which is analogous to the spin rotation of a massive Dirac-type neutrino~\cite{fujikawa}; the left-handed neutrino rotates in a magnetic field to a right-handed neutrino which has no charged current weak interaction, and vice versa. Thus both of the particle and anti-particle in our model can have the charged current weak interaction inside the strong magnetic field; by this way, one can confirm the entangled helicity in our model. This use of weak interaction to analyze the spin freedom of a charged electron at high energies replaces the use of a Stern-Gerlach apparatus~\cite{bohm}. The Stern-Gerlach apparatus is useless for the electron~\cite{tolhoek}: A well-known argument going back to Bohr and Mott~\cite{mott} shows that the inhomogeneity of the magnetic field causes a spreading of a charged electron beam (the particles of the Stern-Gerlach experiment are electrically neutral), which is so large that the spreading arising from different orientations of the magnetic moment in the inhomogeneous magnetic field is not detectable. In passing, we  mention an interesting polarization measurement related to strong interactions~\cite{sakai}.

As for {\em space-time} non-locality and causality, our formulas in (1)-(4) are manifestly Lorentz invariant (if not covariant), and the locality in the Lagrangian level is manifest and no particles propagate faster than the light. Note that the statements made in the center of mass frame of $P$ are Lorentz invariant. It is significant that we integrate over the coordinates and time  of the interaction point in Minkowski space in defining (3). We do not specify when and where the decay of the heavy particle took place. Our formula provides no information about the relative space-time coordinates of the electron and positron which appear in the asymptotic state (8); instead, we specify the energy and momentum of each particle precisely, to be perfectly consistent with the uncertainty principle. Dyson's formula tells us what we can talk about and what we cannot talk about in the S-matrix theory of local relativistic quantum field theory.  The relative space-time coordinates of two particles in the asymptotic state are completely indeterminate (i.e., one cannot tell if they are space-like or time-like) to be consistent with the uncertainty relation, and in this sense the uncertainty principle and the possible non-locality issue associated with entanglement appear to be related. 

As for the consistency with local realism represented by non-contextual and local hidden-variables models~\cite{mermin}, we here mention the condition which does not appear to be widely recognized; this condition states that non-contextual and local hidden-variables models in the Hilbert space with dimension $d=4$ consisting of two spin freedom (such as the model used in the original paper of Bell~\cite{bell}), for example, can describe only separable pure states~\cite{fuji-oh-zhang}. The basic observation involved in deriving this condition is that the non-contextual and local hidden-variables model in $d=4$ is reduced to a factored product of two non-contextual hidden-variables models in $d=2$ if one asks that the hidden variables model gives CHSH inequality 
$|\langle B\rangle| \leq 2$~\cite{chsh, cirel'son} uniquely, as one usually does
in the name of Bell's theorem. This condition is a hidden-variables version of
Gisin's theorem~\cite{werner, gisin} which states that $|\langle B\rangle| \leq 2$ implies pure separable states; it is also consistent with Gleason's theorem~\cite{gleason} which excludes $d=4$ non-contextual hidden-variables models but allows $d=2$ models. Non-contextual and local hidden-variables models then imply the condition~\cite{fuji-oh-zhang} 
\begin{eqnarray}
\Psi(\vec{0})^{\dagger}P_{a}(s^{\prime})\otimes P_{b}(s^{\prime\prime})
\Psi(\vec{0})=\Psi(\vec{0})^{\dagger}P_{a}(s^{\prime})\Psi(\vec{0})\Psi(\vec{0})^{\dagger}P_{b}(s^{\prime\prime})\Psi(\vec{0})
\end{eqnarray}
for any choice of the projection operators $P_{a}(s^{\prime})$ and 
$P_{b}(s^{\prime\prime})$ in (23) for the asymptotic state $\Psi(\vec{0})$
in (17).
It is easy to confirm that this relation does not hold for a suitable choice of $s^{\prime}$ and $s^{\prime\prime}$ such as $s^{\prime}=s^{\prime\prime}=s$, which shows that local realism  is not consistent with the state (17) in agreement with the more conventional analysis using CHSH inequality. 

As for the entanglement of relativistic particles in general, the photon associated with the electromagnetic field is massless and always ultra-relativistic but it has different characteristics~\cite{weinberg,caban2}. The electromagnetic field is a vector field $A_{\mu}(x)$ and can be described not only by the circular polarization (helicity) but also by the "transverse linear polarization", which is directly measured, namely, to stay linearly polarized before and after the measurement, and thus characterized by a continuous number of projection operators which are absent in the case of the spin $1/2$ electron described by the operator (15) in the ultra-relativistic limit. From the point of view of a modified Dirac spin operator, the suppression of the longitudinal linear polarization of the photon is analogous to the Lorentz suppression of transverse spin components of the ultra-relativistic electron in (15).  The possible suppression of entanglement measured by a modified Dirac spin operator in the ultra-relativistic limit we argued does not happen for the entanglement of the electromagnetic field, which is described by the 2-photon decay amplitude of the form 
\begin{eqnarray}
\Psi=\int d^{4}x P(x)\vec{E}(x)\cdot\vec{B}(x)|P(\vec{0})\rangle,
\end{eqnarray}
if one uses the linear polarization. 
\\

Finally, we mention the closely related works on spin operators in the past. The spin operators using the Pauli-Lubanski vector both in relativistic quantum mechanics and in the first quantization level of Dirac equation have been proposed in~\cite{czachor}. The transverse components of spin operators thus defined are shown to be suppressed in the ultra-relativistic limit, and this was illustrated for an entangled pair of two fermions (not a  fermion-antifermion pair) propagating together in the {\em same} direction; this was one of the early and pioneering treatments of relativistic entanglement.  Physically, this process is quite different from our example of the asymptotic state (8); one can always choose the rest frame of two fermions  by a suitable Lorentz transformation in the example of~\cite{czachor} (namely, the Lorentz frame dependence of entanglement is analyzed there),  while we cannot choose a Lorentz frame where both of the electron and positron are at rest for the decay of a very massive particle and, in fact, we analyze the intrinsic Lorentz frame independent property.  The treatment of an anti-particle is not given in the relativistic quantum mechanics nor the linear polarization of the photon is mentioned in~\cite{czachor}. In the first quantization, the notion of chirality is not explicit and the distinction between the spin operator and the magnetic moment is not obvious. We know the correct theory of leptons and quarks, namely, the Standard Model, and our field theoretical construction  of states and spin operators is straightforward and simpler without referring to the Pauli-Lubanski vector. Our view is that the Standard Model
can describe all the observable properties of leptons.

An elaborate construction of a spin operator using Pauli-Lubanski vector 
is given in~\cite{caban1}, but the final outcome is essentially the same as our spin operator in (13) obtained from the Noether current associated with rotational symmetry of the free Dirac action. The decay of a massive pseudo-scalar is then analyzed, but our construction of the final asymptotic state of the decay of a massive pseudo-scalar particle by Dyson's formula is simpler, which takes care of both electron and positron without any obvious contradiction with locality, causality and the uncertainty principle. One can confirm that the spin correlation in Ref.~\cite{caban1} is obtained from the spin operator (13) (Wigner spin) and the state (8) in the present paper. 
Physically, the basic difference between the modified Dirac spin operator we analyzed in detail and that in~\cite{caban1} is whether we can measure the well-defined transverse polarization of an ultra-relativistic free fermion at the {\em laboratory frame}. 

We comment on the  recent stimulating papers~\cite{vedral1} which criticize the use of  most conventional spin operators. In~\cite{vedral1}, the authors argue that the Pauli-Lubanski (or similar) spin operators, which are related to our operator in (15), are not suitable to describe measurements where spin couples to an electromagnetic field in the measuring apparatus. This assertion is partly related to the fact that the distinction between the spin angular momentum and the magnetic moment is not transparent in the first quantization they use. In reality, a  helicity eigenstate is represented as a linear superposition of transversely polarized states when the fermion with 
non-vanishing magnetic moment enters a transverse magnetic field~\cite{fujikawa}. 
The Gordon decomposition~\cite{bjorken} may be useful in the analysis of electromagnetic interactions of charged particles. In this connection, the experimental review paper~\cite{tolhoek} which clearly distinguishes the spin angular momentum and the magnetic moment of the high energy electron is valuable; this reference  emphasizes the fact that a Stern-Gerlach apparatus, on which the argument of~\cite{vedral1} is based, is useless for the electron we are interested in. The weak interaction we discussed can in principle measure the helicity of elementary particles directly. 

We also mention that the critical reassessment of various definitions of spin operators in the first quantization with an emphasis on the Foldy-Wouthuysen mean-spin operator~\cite{foldy}  has been recently given both from a theoretical point of view~\cite{pcaban} and from a measurement point of view~\cite{bauke}.
\\
   
\noindent {\bf Acknowledgments}
\\

\noindent 
We thank Sakue Yamada for a helpful comment on the polarization of the electron in a high energy synchrotron. We also thank Sixia Yu for numerous helpful comments.
One of the authors (KF) thanks the hospitality at the Center for Quantum Technologies, National University of Singapore. This work is partially supported by the National Research Foundation and Ministry of Education, Singapore (Grant No. WBS: R-710-000-008-271), and JSPS KAKENHI (Grant Number 25400415).

\end{document}